\documentclass[twocolumn,prb,showpacs]{revtex4}
\usepackage{graphicx,color}

\begin{document}


\title{
Superconductivity induced by ruthenium substitution in an iron arsenide:
investigation of SrFe$_{2-x}$Ru$_x$As$_2$ ($0 \leq x \leq 2$)
}

\author{W.\ Schnelle}
\author{A.\ Leithe-Jasper}
\author{R.\ Gumeniuk}
\author{U.\ Burkhardt}
\author{D.\ Kasinathan}
\author{H.\ Rosner}

\affiliation{Max-Planck-Institut f\"ur Chemische Physik fester Stoffe,
N\"othnitzer Stra{\ss}e 40, 01187 Dresden, Germany}

\begin{abstract}
The magnetism in SrFe$_2$As$_2$ can be suppressed by electron
doping through a small substitution of Fe by Co or Ni, giving
way to superconductivity. We demonstrate that a massive
substitution of Fe by isovalent ruthenium similarly suppresses
the magnetic ordering in SrFe$_{2-x}$Ru$_x$As$_2$ and leads to
bulk superconductivity for $0.6 \leq x \leq 0.8$. Magnetization,
electrical resistivity, and specific heat data show $T_c$ up to
$\approx 20$\,K. Detailed structural investigations reveal a
strong decrease of the lattice parameter ratio $c/a$ with
increasing $x$. DFT band structure calculations are in line with
the observation that the magnetic order in
SrFe$_{2-x}$Ru$_x$As$_2$ is only destabilized for large $x$.
\end{abstract}


\maketitle

\section{Introduction}

Soon after the discovery of superconductivity (SC) in doped
$R$FeAsO ($R$ = rare-earth element) materials,\cite{Kamihara08a}
investigations also focused on the structurally related
$A$Fe$_2$As$_2$ compounds ($A$ = alkaline, alkaline-earth, or
rare-earth metal). In the latter structures, the Fe$_2$As$_2$
slabs are separated only by single elemental $A$
layers.\cite{Rotter08a} The compounds become superconductors if
appropriately modified by substitutions on the $A$ site by
alkali metals\cite{Rotter08b,Sasmal08a,GFChen08a} or direct
substitution within the Fe$_2$As$_2$ slab by
Co\cite{LeitheJasper08b,Sefat08b,CWang08a,Kumar08b} or
Ni.\cite{LJLi08a} Recently, also the appearance of SC upon
substitution of As by P was reported.\cite{SJiang09a} Controlled
tuning of the electronic structure by selective substitutions
provides an opportunity to test and refine theoretical models
since these substitutions can introduce charge carriers, modify
the lattice parameters, and may significantly suppress the
structural/magnetic transitions observed in the ternary parent
compounds.\cite{Rotter08a}

Application of external pressure has been understood as a
``clean'' alternative to substitutions in tuning the electronic
state. $A$Fe$_2$As$_2$ compounds indeed show crossover to SC at
pressures as low as 0.4 GPa for
CaFe$_2$As$_2$,\cite{HLee08a,Park08a,Torikachvili08a} and
high $T_c$ were observed (27\,K at 3\,GPa for SrFe$_2$As$_2$;
29\,K at 3.5\,GPa for BaFe$_2$As$_2$).\cite{Alireza08a}
Pressure reduces the antiferromagnetic (AFM) phase transition
temperature ($T_0$) in SrFe$_2$As$_2$ and an abrupt loss of
resistivity hints for the onset of
SC.\cite{Kumar08a,Igawa08a,Torikachvili08b} However, the
nature of pressure (hydrostatic vs.\ anisotropic strains) in
these experimental procedures is currently up for
debate\cite{WYu08a}.

A feature common to both approaches (chemical substitution or
external pressure) is the correlation of anisotropic changes in
the crystal lattice with the suppression of the spin density
wave (SDW) type of AFM. Transition-metal ($T$) substitution
studies in $A$Fe$_{2-x}T_x$As$_2$ carried out so far show a
significant contraction of the tetragonal $c$ axis
length\cite{LeitheJasper08b,Sefat08b,LJLi08a} and SC upon
partial substitution of Fe by Co or Ni (electron doping) and an
opposite trend if Fe is replaced by Mn (hole
doping).\cite{Kasinathan09a} For the latter substitution
series the SDW transition temperature $T_0$ is not suppressed
with increasing $x$ and no SC is
observed.\cite{Kasinathan09a} In contrast, in
indirectly-doped Sr$_{1-x}$K$_x$Fe$_2$As$_2$ and
Ba$_{1-x}$K$_x$Fe$_2$As$_2$\cite{Sasmal08a,GFChen08a,Rotter08b}
the $a$ lattice parameter decreases with $x$ while $c$
increases, keeping the unit cell volume almost constant. For
BaFe$_2$As$_{2-x}$P$_x$, an isovalent substitution where SC is
observed for $x > 0.5$, both $a$ and $c$ decrease with
$x$.\cite{SJiang09a} The fact that substitutions within the
Fe$_2$As$_2$ slab with other $d$-metals lead to
superconductivity, albeit with lower $T_c$ than for indirect
doping, favors an itinerant electronic theory, in contrast to
the strongly correlated cuprates.\cite{LeitheJasper08b}

A point to note is that the modification of the electron count
by substitutions is inevitably connected with structural changes
in the Fe$_2$As$_2$ slabs. Unfortunately, for most substitution
series only the unit cell dimensions are reported while further
crystallographic data (bonding angles) are unknown. For pressure
studies on Sr/BaFe$_2$As$_2$ compounds aiming at physical
properties it is difficult to connect the results to variations
of the crystal structure since corresponding compressibility
studies are largely missing. Thus, currently, it is not known
how the valence electron concentration \textit{and/or} the
structural parameters have to be modified by substitution in
order to suppress the magnetism and eventually generate SC in
$A$Fe$_2$As$_2$ compounds.

Recently, Nath \textit{et al}.\cite{Nath09a} reported on the
physical properties of SrRu$_2$As$_2$ and BaRu$_2$As$_2$, which
are isostructural to SrFe$_2$As$_2$ and do not show
SC.\cite{Jeitschko87a} On the other hand, the isostructural
LaRu$_2$P$_2$ is a long-known superconductor with $T_c$ =
4.1\,K.\cite{Jeitschko87a} In this communication we present
results of a study of the solid solution
SrFe$_{2-x}$Ru$_x$As$_2$. A massive substitution of Fe by
nominally isoelectronic Ru suppresses the SDW-ordered state and
bulk superconductivity is observed for $0.6 \leq x \leq 0.8$.
Characterization of the SC state by magnetic susceptibility,
electrical resistivity, and specific heat measurements shows a
maximum $T_c$ of $\approx 20$\,K. Detailed crystallographic
information of all samples of the series are obtained from full
profile refinement of powder X-ray diffraction data. Band
structure calculations confirm that AFM order is only suppressed
for \textit{large} $x$, giving way to SC in
SrFe$_{2-x}$Ru$_x$As$_2$.

In this study we obtain the pertinent crystallographic details
for a complete substitution series with magnetically ordered,
superconducting, and normal-metallic ground states. Such data
are a prerequisite for a future comparative study of
substitution systems (with or without superconducting phases)
based on the SrFe$_2$As$_2$ parent compound. Another aspect is
that substitution with nominally isovalent Ru -- in contrast to
the supposed electron doping by Co or Ni -- is expected not to
change the charge. In this way also no charge disorder is
generated within the Fe$_2$As$_2$ slab. This is in contrast to
substitutions with $d$ metals from the cobalt or nickel group.
On the other hand, Ru substitution possibly will generate
disorder due to its heavier mass and larger size. Both
scattering mechanisms seem to limit the achievable $T_c$ of
materials with in-plane substitutions, in contrast to SC
material obtained through indirect doping of the Fe$_2$As$_2$
slabs by substitution on the $A$ site.

\section{Experimental and crystal structure}

Samples were prepared by powder metallurgical techniques.
Blended and compacted mixtures of precursor alloys SrAs,
Fe$_2$As together with As and Ru powder were placed in
glassy-carbon crucibles, welded into tantalum containers, and
sealed into evacuated quartz tubes for heat treatment at
900\,$^\circ$C for 24\,h to 7\,d followed by several regrinding
and densification steps. Samples were obtained in the form of
sintered pellets. Details of powder XRD procedures and
electron-probe microanalysis (EPMA) are given in previous
publications.\cite{LeitheJasper08b,Kasinathan09a} The
magnetic susceptibility was measured in a SQUID magnetometer
(MPMS) and heat capacity by a relaxation method (PPMS, Quantum
Design). The electrical resistivity was determined by a
four-point dc method (current density $<3$\,A\,mm$^{-2}$). Due
to the contact geometry the absolute resistivity could be
determined only with an inaccuracy of $\pm$30\,{\%}.

Band structure calculations were performed within the local
density approximation (LDA) using the full potential local
orbital code FPLO (v.\ 8.00) with a $k$-mesh of
24$\times$24$\times$24 $k$-points and the Perdew-Wang
parameterization of the exchange-correlation potential. The used
structural parameters were those from Table\ \ref{thetable}.
\cite{FeConote}

\section{Results and discussion}

\begin{table*}[htb]
\begin{center}
\caption{Crystallographic and electronic data for
SrFe$_{2-x}$Ru$_x$As$_2$: lattice parameters $a$, $c$, ratio
$c/a$, cell volume $V$, refined (by WinCSD\cite{WinCSD})
positional parameter $z$ of As, interatomic distance
$d_\mathrm{T-As}$, refined Ru content $x$, intensity ($R_I$) and
profile ($R_P$) residuals, SDW ($T_0$) and superconducting
transition temperature ($T_c^\mathrm{mag}$; crossing of the
tangent to the upper part of the fc susceptibility transition
with $\chi$ = 0) and from specific heat ($T_c^\mathrm{cal}$;
from two fluid model fit, see text), idealized specific heat
jump $\Delta c_p/T_c^\mathrm{cal}$, linear term $\gamma$ to
$c_p$, Debye temperature calculated from $\beta$. $T_c$ are only
listed for bulk superconductors. fil. = observation of
superconducting traces (filaments).
\label{thetable}}
\begin{ruledtabular}
\begin{tabular}{l|cccccccc|cccccc}
$x$ & $a$        & $c$        & $c/a$  & $V$       & $z_\mathrm{As}$ & $d_\mathrm{T-As}$ & $x$  & $R_I$/$R_P$ & $T_0$        & $T_c^\mathrm{mag}$ & $T_c^\mathrm{cal}$ & $\Delta c_p/T_c^\mathrm{cal}$ & $\gamma$            & $\Theta_D$          \\
nom.& [\AA]      & [\AA]      &        & [\AA$^3$] &                 & [\AA]                   & ref. & {\%}        & [K]          & [K]                & [K]                & \multicolumn{2}{c}{[mJ/mol\,K$^2$]}                 & [K]           \\ \colrule
0.0\footnotemark[1]
    & 3.9243(1)  & 12.3644(1) & 3.1507 & 190.5     & 0.3600(1)       & 2.388(1)          & 0    & -      -    & 203          & -                  & -                  & -                             & -                   & -                   \\
0.1 & 3.93210(3) & 12.3446(1) & 3.1347 & 190.9     & 0.3603(1)       & 2.3911(3)         & 0.12 & 3.7/6.3     & 190          & -                  & -                  & -                             & -                   & -                   \\
0.2 & 3.94387(2) & 12.2905(1) & 3.1125 & 191.2     & 0.3602(1)       & 2.3922(3)         & 0.21 & 3.6/6.8     & 165          & -                  & -                  & -                             & -                   & -                   \\
0.3 & 3.95145(3) & 12.2551(2) & 3.1014 & 191.4     & 0.3601(1)       & 2.3924(4)         & 0.31 & 3.8/10.3    & $\approx$140 & -                  & -                  & -                             & -                   & -                   \\
0.4 & 3.96689(2) & 12.1772(1) & 3.0697 & 191.6     & 0.3599(1)       & 2.3925(4)         & 0.42 & 4.6/8.4     & $\approx$100 & fil.               & -                  & -                             & -                   & -                   \\
0.5 & 3.97720(4) & 12.1300(2) & 3.0499 & 191.9     & 0.3597(1)       & 2.3927(5)         & 0.50 & 5.3/14.1    & -            & fil.               & $<$2.0             & -                             & -                   & -                   \\
0.6 & 3.99178(2) & 12.0635(1) & 3.0221 & 192.2     & 0.3599(1)       & 2.3959(4)         & 0.61 & 4.2/7.8     & -            & 19.3               & 19.8               & 13.4                          & 6.2                 & 232                 \\
0.7 & 4.00507(2) & 12.0087(1) & 2.9983 & 192.6     & 0.3598(1)       & 2.3976(3)         & 0.71 & 3.3/6.3     & -            & 19.3               & 20.1               & 11.6                          & 7.3                 & 229                 \\
0.8 & 4.01096(2) & 11.9835(1) & 2.9877 & 192.8     & 0.3598(1)       & 2.3983(4)         & 0.81 & 4.9/7.5     & -            & 17.6               & 17.2               & 13.6                          & 6.7                 & 231                 \\
1.0 & 4.04437(6) & 11.8097(3) & 2.9200 & 193.2     & 0.3597(1)       & 2.4015(5)         & 1.03 & 4.5/10.2    & -            & fil.               & $<$2.0             & 0                             & 6.9                 & 243                 \\
1.5 & 4.09818(2) & 11.5301(1) & 2.8135 & 193.6     & 0.3593(1)       & 2.4056(4)         & 1.50 & 3.8/7.4     & -            & -                  & -                  & -                             & -                   & -                   \\
2.0\footnotemark[2]\footnotemark[3]
    & 4.16911(2) & 11.1706(1) & 2.6794 & 194.2     & 0.3591(1)       & 2.4148(4)         & 2    & 4.2/7.6     & -            & -                  & -                  & -                             & 4.1\footnotemark[2] & 270\footnotemark[2] \\
\end{tabular}
\mbox{\footnotetext[1]{from Ref.\ \onlinecite{Tegel08a}}
\footnotetext[2]{$a$ = 4.1713\,\AA, $c$ = 11.1845\,\AA, $V$ = 194.6\,\AA$^3$ (from Ref.\ \onlinecite{Nath09a})}
\footnotetext[3]{$a$ = 4.168\,\AA,  $c$ = 11.179\,\AA,  $V$ = 194.2\,\AA$^3$ (from Ref.\ \onlinecite{Jeitschko87a})}}
\end{ruledtabular}
\end{center}
\end{table*}
\noindent

For all samples the crystal structure of ThCr$_2$Si$_2$ type
(space group $I4/mmm$)\cite{Tegel08a} was refined from powder
XRD data by full profile methods (Sr in 2$a$ (0, 0, 0), Fe/Ru in
4$d$ (0, 1/2, 1/4), As in 4$e$ (0, 0, $z$)) (see Table
\ref{thetable}). In the powder X-ray diffractograms of samples
with $x \geq 0.5$ broadening of (00$l$) reflections are observed
suggesting some local disorder along [001]. Nevertheless, they
give no evidence for superstructure formation due to long-range
Ru ordering. The refined lattice parameters, the refined Ru
occupancies, as well as EPMA unambiguously reveal the
substitution of Fe by Ru. The nominal Ru contents are in good
agreement with both the Ru occupancies from XRD and the EPMA
data. The samples contained as minor impurity
Ru$_{1-x}$Fe$_x$As. Upon exchange of Fe by Ru a strong linear
decrease of the $c$ parameter of the unit cell is observed
($-$9.7\,{\%} for $x = 2$). The tetragonal $a$,$b$ plane and
with it the transition-metal distance $d_{T-T} = a/\sqrt{2}$
expands by the substitution (by +6.2\,{\%} for $x = 2$). The
increase of the distances $d_{T-\mathrm{As}}$ with $x$ on the
other hand is small (+1.1\,{\%} for $x = 2$) and the $z$
parameter of As decreases by only 0.25\,{\%}. Surprisingly, the
exchange of Fe by the larger Ru atoms\cite{Emsley92} results
only in a 2.0\,{\%} increase in the unit cell volume $V$. The by
far largest structural effect of Ru (or
Co)\cite{LeitheJasper08b} substitution is the strong decrease of
the $c/a$ ratio, i.e.\ a strong strain-like deformation with
respect to the crystal lattice of the ternary Fe parent
compound. Correspondingly, the tetrahedral bonding angles
$\epsilon_{1,2}$ As--(Fe,Ru)--As depart from each other with
increasing $x$. There are no visible discontinuous changes in
these room-temperature structure data which could be connected
to the various electronic ground state of
SrFe$_{2-x}$Ru$_x$As$_2$.

\begin{figure}[htb]
\includegraphics[height=3.4in,angle=90]{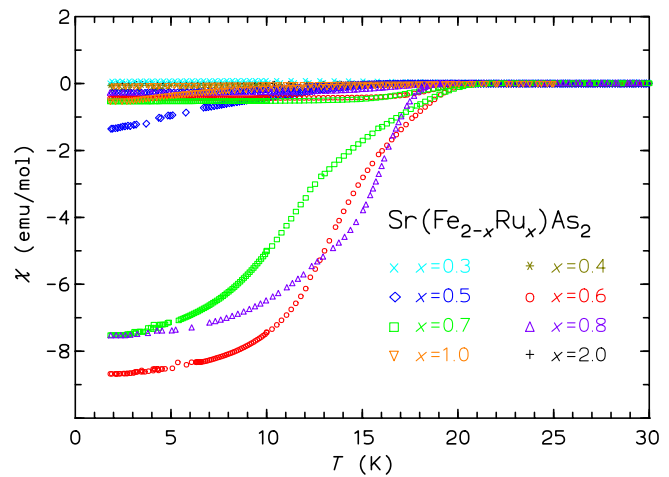}
\caption{(Color online)
Magnetic susceptibility $\chi(T)$ of SrFe$_{2-x}$Ru$_x$As$_2$
samples in a magnetic field $\mu_0H$ = 2\,mT.
\label{figchilo}}
\end{figure}

\begin{figure}[htb]
\includegraphics[height=3.4in,angle=90]{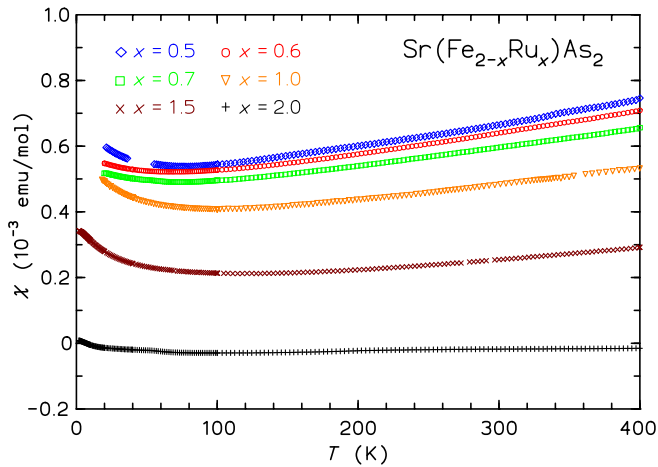}
\caption{(Color online)
Corrected high-field magnetic susceptibility $\chi(T)$ of
selected SrFe$_{2-x}$Ru$_x$As$_2$ samples.
\label{figchihi}}
\end{figure}

In Fig.\ \ref{figchilo} the low-field magnetic susceptibility of
Sr(Fe$_{2-x}$Ru$_x$)As$_2$ samples is plotted. Weak diamagnetic
signals for $T < 16$\,K in warming after zero-field cooling
(zfc) are already visible for a Ru concentration $x = 0.4$.
However, the diamagnetic shielding signal increases dramatically
from $x = 0.5$ to $x = 0.6$. For $x = 0.6$, 0.7, and 0.8 the
shielding comprises the full volume of the sample suggesting
bulk SC. On the other hand, the Meissner effect
(measured during field cooling) is extremely small. This
peculiarity is also observed for other $A$(Fe$_{2-x}T_x$)As$_2$
materials\cite{LeitheJasper08b,Sefat08b} and is probably due to
strong pinning in materials with substitutions on the iron site,
i.e.\ within the superconducting slab. The sample with $x =
1.0$ also does not show bulk SC but only a very
small diamagnetism after zfc. The superconducting transition
temperatures are listed in Table \ref{thetable}, however the
transitions are rather broad. This observation of diamagnetic
traces and the appearance of superconducting filaments for low
Ru concentrations is certainly due to microscopic
inhomogeneities of the Ru distribution.

The high-temperature susceptibility (high-field data eventually
corrected for small ferromagnetic impurities, Fig.\
\ref{figchihi}) of SrFe$_{2-x}$Ru$_x$As$_2$ with $x \leq 1.5$ is
generally paramagnetic and shows a typical linear increase with
$T$ for $T > 100$\,K, similar to that of compounds of the Co
substituted system.\cite{LeitheJasper08b,XFWang09a} The absolute
values decrease systematically with the Ru content $x$ as does
the linear $T$ dependence. SrRu$_2$As$_2$ finally is diamagnetic
and shows no significant slope of $\chi(T)$ above 100\,K, in
agreement with Ref.\ \onlinecite{Nath09a}. No phase transitions
(except for SC) are observed for $x > 0.5$. The samples with $x
\leq 0.4$ display anomalies at around 200\,K, 190\,K, and
150--170\,K for $x$ = 0, 0.1, and 0.2, respectively. These
temperatures are close to the temperatures $T_0$ in Table
\ref{thetable} which were identified from the anomalies in the
resistivity $\rho(T)$ (see below) which mark the SDW transition.

\begin{figure}[htb]
\includegraphics[height=3.4in,angle=90]{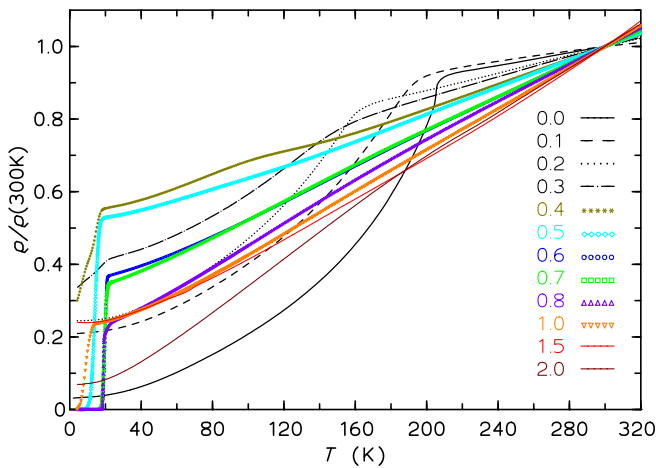}
\caption{(Color online) Electrical resistivity (normalized at
300\,K) of SrFe$_{2-x}$Ru$_x$As$_2$ samples ($x$ as indicated on
the curves). Data for $x = 0$ from Ref.\
\onlinecite{Krellner08a}.
\label{figrho}}
\end{figure}

\begin{figure}[htb]
\includegraphics[width=3.4in,angle=0]{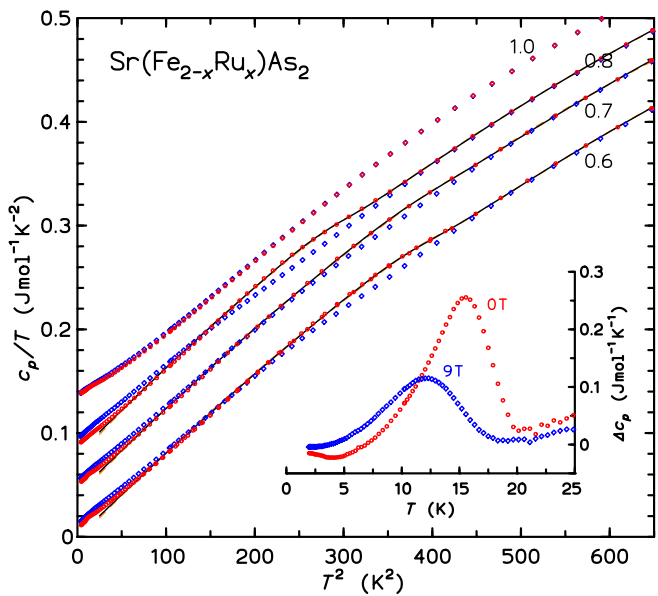}
\caption{(Color online) Specific heat $c_p/T$ vs.\ $T^2$ of
SrFe$_{2-x}$Ru$_x$As$_2$ ($x$ = 0.6, 0.7, 0.8, 1.0). Data for
zero magnetic field (red circles) are shown with the
corresponding two-fluid (full black line) and BCS (orange dashed
line) type fits (see text). For a field $\mu_0H = 9$\,T only
data (blue diamonds) are shown. Data for $x$ = 0.7, 0.8, 1.0 are
shifted by 0.04, 0.08, and 0.12 units upwards, respectively.
Inset: difference of the specific heats ($\Delta c_p$) of the SC
sample with $x = 0.8$ and the sample $x = 1.0$ (no bulk SC).
Both the difference curves for zero and 9\,T fields are given.
\label{figcp}}
\end{figure}

Due to the inaccuracy of the contact geometry we prefer to plot
normalized electrical resistivity in Fig.\ \ref{figrho}. The
room temperature resistivity values $\rho$(300\,K) are 500--800
$\mu\Omega$\,cm for samples with $x = 0$ and low Ru content $x$.
The $\rho$(300\,K) values decrease slightly with increasing $x$
and for SrRu$_2$As$_2$ we find $\rho$(300\,K) of only $\approx$
230 $\mu\Omega$\,cm. The temperature dependence of
$\rho(T)/\rho$(300\,K) corroborates the superconducting
transitions. Zero resistivity is observed for Ru concentrations
of $x = 0.5$, 0.6, 0.7, and 0.8 at 7.0\,K, 17.5\,K, 16.0\,K, and
18.0\,K, respectively. The sample with $x = 1.0$ also shows a
drop in $\rho(T)$ below 18\,K, however $\rho = 0$ is not reached
at 4\,K, indicating only filamentary SC. Also, a small drop in
$\rho(T)$ at low $T$ is already visible for $x = 0.4$.

The resistivity of the sample $x$ = 0.4 however also shows a
kink at $T_0 \approx 100$\,K, the $x = 0.3$ sample at $\approx
140$\,K. For lower Ru concentration $T_0$ increases continuously
(see Table \ref{thetable}). The kink in $\rho(T)$, which roughly
coincides with anomalies in the high-field susceptibility (see
above), is the signature of the SDW
transition\cite{Tegel08a,Krellner08a,LeitheJasper08b,Kasinathan09a}
which is found at $T_0 = 203$\,K in
SrFe$_2$As$_2$.\cite{Tegel08a} Interestingly, in
Sr/BaFe$_{2-x}$Co$_x$As$_2$ crystals the resistivity for all $x
> 0$ increases below $T_0$ while for unsubstituted
Sr/BaFe$_2$As$_2$ $\rho(T)$ decreases below $T_0$ (see, e.g.,
Refs.\ \onlinecite{LeitheJasper08b,Ahilan09a}). In contrast, for
SrFe$_{2-x}$Ru$_x$As$_2$ the resistivity \textit{decreases}
below $T_0$ for all concentrations of Ru. Generally, for a full
opening of a gap due to an SDW ordering an increase of $\rho(T)$
would be expected. Instead, it seems that the behavior of
$\rho(T)$ below $T_0$ is connected to the presence or absence of
charge disorder caused by electron doping. Obviously, for
SrFe$_{2-x}$Ru$_x$As$_2$ and undoped materials, in absence of
such disorder the mechanism leading to a decrease of $\rho(T)$
below $T_0$ is dominating.

The specific heat of selected samples is given in Fig.\
\ref{figcp}. Clear albeit broadened anomalies are seen close the
transitions temperatures $T_c^\mathrm{mag}$ indicated by the
low-field magnetization (see table \ref{thetable}). The sample
with $x = 0.7$ shows a less pronounced transition than the
samples with neighboring compositions. In agreement with the
reduced resistive $T_c$ we conclude that this sample is
inhomogeneous and of lower quality than the other samples. The
sample with $x = 0.8$ displays a pronounced transition at
17.2\,K. The composition SrFeRuAs$_2$ shows no bulk SC above
2\,K. Since the specific heats of the two latter samples are
very similar for $T > 20$\,K the data for $x = 1$ may serve as a
reference for phonon and normal electronic contributions to the
$x = 0.8$ data. The inset in Fig.\ \ref{figcp} shows the
difference of the specific heat of the samples. The size of the
resulting step $\Delta c_p/T_c$ as evaluated by the usual
entropy-conserving construction (equal areas in $c_p/T$) for $H$
= 0 is 18.4(1.8) mJ\,mol$^{-1}$ K$^{-2}$ and $T_c$ = 17.3(5)\,K.

In order to obtain further electronic and phononic properties
from the specific heat, the data between 5.0\,K and 25.5\,K were
fitted with a model including a phonon contribution (harmonic
lattice approach $c_\mathrm{ph} = \beta T^3 + \delta T^5$) and
an electronic term according to the weak coupling BCS theory or
the phenomenological two-fluid model. The latter model is a good
approximation for the thermodynamic properties of some strong
coupling superconductors. Folding with a Gaussian simulates the
broadening of the transitions due to chemical inhomogeneities.
The parameters resulting from the fits are given in Table
\ref{thetable}. The relative specific heat step $\Delta c_p/T_c$
at $T_c^\mathrm{cal}$ is quite similar for the three
investigated SC samples. It has to be remarked that these values
are smaller than the value obtained from the difference of the
samples with $x = 0.8$ and $x = 1.0$. Nevertheless, the size of
the specific heat step at $T_c$ is small and comparable to
values observed for superconducting compositions of
SrFe$_{2-x}$Co$_x$As$_2$ (10--13 mJ\,mol$^{-1}$K$^{-2}$; hole
doping within the Fe$_2$As$_2$ slab),\cite{LeitheJasper08b} but
much smaller than for, e.g., Ba$_{0.6}$K$_{0.4}$Fe$_2$As$_2$
($\approx$ 98 mJ\,mol$^{-1}$K$^{-2}$).\cite{GMu08a} The fits
with the two-fluid models (shown in Fig.\ \ref{figcp}) are
superior to those with the BCS model. The ratio $\Delta
c_p/(\gamma T_c^\mathrm{cal})$ is $\approx 2.1$, significantly
larger than the weak coupling BCS limit (1.43). Both these
findings indicate strongly coupled SC, in agreement with $\mu$SR
measurements (see, e.g., Ref.\ \onlinecite{Khasanov09a}).

\begin{figure}[t]
\includegraphics[width=3.4in,angle=0]{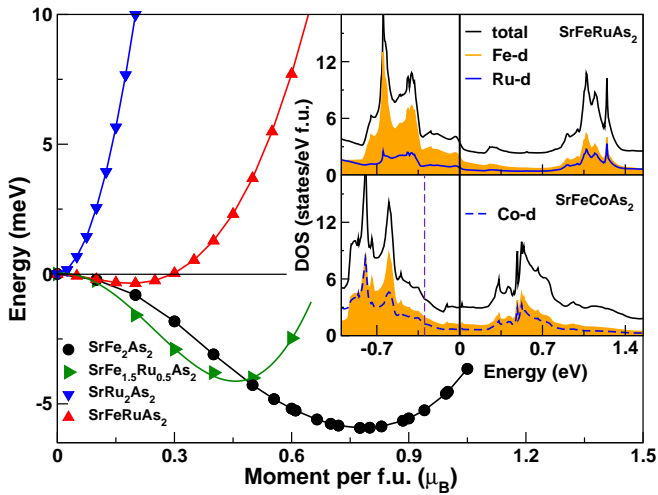}
\caption{(Color online) 
FSM curves for SrFe$_{2-x}$Ru$_x$As$_2$ for $x$ = 0, 0.5, 1 and
2. The inset shows a comparison of the total and partial DOS for
SrFeRuAs$_2$ and SrFeCoAs$_2$. The dashed perpendicular line
shows the Fermi level for SrFe$_2$As$_2$.
\label{figfsm}}
\end{figure}

To understand the changes in the electronic structure upon
substitution of Fe by Ru, we carried out band structure
calculations for $x$ = 0, 0.5, 1, 2. The partial Ru substitution
was modeled by supercells. Since any SDW pattern would be
strongly influenced by the choice of the particular supercell,
especially for the larger Ru content, we decided to compare the
stability with respect to magnetism applying the fixed spin
moment (FSM) approach (cf.\ Fig.\ 3 in Ref.\
\onlinecite{Krellner08a}). The resulting curves are shown in
Fig.\ \ref{figfsm}. As one may expect, the magnetic moment and
the energy gain due to magnetic ordering are reduced with
increasing Ru content $x$. The destabilization is especially
strong between $0.5 \le x \le 1$, the fully Ru substituted
compound is clearly nonmagnetic. In accord with the classical
Stoner picture for itinerant magnets, the suppression of
magnetism originates from a strong decrease of the density of
states (DOS) at the Fermi level $N(\varepsilon_F)$ = 4.9, 4.4,
3.5 and 1.9 states/(eV f.u.) for $x$ = 0, 0.5, 1, 2, in
agreement with experimental $\gamma$ values (Table
\ref{thetable}).

In the real system, the magnetism will be further destabilized
due to the Fe-Ru disorder. In contrast to the Co substituted
compound, where the Co and Fe 3$d$ states are almost
indistinguishable (see inset of Fig.\ \ref{figfsm}), the Ru 4$d$
states differ considerably in the region close to
$\varepsilon_F$. This difference has its origin in the different
potential and lower site energy of Ru with respect to Fe. In
turn, the distinct potential will be responsible for an enhanced
scattering and therefore destabilize magnetic ordering. In the
Co substituted compound, however, only the shift of the Fermi
level due to the additional electron (see Fig.\ \ref{figfsm})
leads to a nonmagnetic state.

Although our calculation illustrate semi-quantitatively the
suppression of magnetism (clearing the scene for incipient SC)
in SrFe$_{2-x}$Ru$_x$As$_2$, the real interplay between
magnetism and the slight volume expansion for increasing Ru
content, accompanied by a reduction of the $c$ axis and small
changes of the As $z$ position is rather complex: the volume
expansion leads to narrower Fe 3$d$ bands that would stabilize
magnetic order, but due to the $c$ axis contraction these states
are simultaneously shifted to lower energy resulting in a
reduced DOS at $\varepsilon_F$. However, the Fe 3$d$ states are
pushed up in energy upon substitution of Ru due to the lower
site energy of the Ru 4$d$ states, compensating the reduction of
$N(\varepsilon_F)$ partially. Furthermore, the Stoner factor $I$
for Ru is significantly smaller than for Fe, disfavoring
magnetic order further.

\section{Discussion and conclusions}

To sum up, we have shown that the partial isovalent
substitution of Ru for Fe in SrFe$_2$As$_2$ suppresses the SDW
transition and gives rise to bulk superconductivity. Both end
members of the series, SrFe$_2$As$_2$ as well as SrRu$_2$As$_2$,
crystallize in the ThCr$_2$Si$_2$ type structure ($I4/mmm$) and
there is no indication of a change or of a superstructure for
intermediate Ru concentrations at the applied experimental
conditions. This observation indicates that Fe and Ru are
isovalent. The lattice parameters $a$ and $c$ vary linearly but
in opposite directions with $x$. As seen from the ratio $c/a$
(decreases by 15\,{\%} for $x$ between 0 and 2) a strong
anisotropic modification of the lattice (a compression of the
tetragonal cell along $c$) is introduced by the exchange of Fe
by Ru. While $z_\mathrm{As}$ does not change significantly with
$x$ this implies a large change of the two different As--Fe--As
bonding angles. While for SrFe$_2$As$_2$ these angles are quite
similar (110.50 and 108.90\,$^\circ$) they deviate quite
strongly in SrRu$_2$As$_2$ (119.43 and
104.73\,$^\circ$).\cite{Tegel08a} Bulk superconductivity
exists in this series for a $c/a$ ratio (at room temperature)
around 3. Electronic properties like the high-temperature
susceptibility also show a smooth variation with $x$.

Whether any of these structural modifications is of special
importance for the occurrence of superconductivity in
SrFe$_{2-x}T_{x}$As$_2$-type substitution series remains to be
determined. Currently, it can be stated that electron doping by
Co (or Ni) is much more efficient in order to suppress the SDW
and allowing for a superconducting ground state than a
compression of the cell along $c$. The fact that Ru
substitutions do not suppress the SDW state for as low
concentrations as for Co, strongly indicates that Ru is
isovalent to Fe in SrFe$_{2-x}$Ru$_x$As$_2$. In (optimally
superconducting) SrFe$_{1.80}$Co$_{0.20}$As$_2$ the $c/a$ ratio
is still quite large ($c/a$ = 3.132), however 0.2 electrons were
added per formula unit.

In order to compare all these parameters further detailed
structural studies on various SrFe$_{2-x}T_x$As$_2$ systems are
required. For the BaFe$_{2-x}$(Co,Ni,Cu)$_x$As$_2$ systems a
first attempt in this direction was presented
recently.\cite{Canfield09a} Also, it is desirable to measure
the compressibility of the compounds under hydrostatic pressure.
Only such data would allow a comparison of pressure experiments
with chemical substitutions studies. During revision of this
work, superconductivity in several systems
Sr/BaFe$_{2-x}T_x$As$_2$ with $T$ =
Ru,\cite{Paulraj09a,YQi09a} Rh,\cite{FHan09b}
Pd,\cite{XZhu09a}, and Ir\cite{FHan09a} was reported. For
some of these systems the electronic state was already addressed
by DFT methods.\cite{LJZhang09a} Experimentally, it appears that
the critical temperatures are generally limited to $\approx
20$\,K and that only for isovalent substitution (\textit{viz.}
Ru) the required level of substitution is such high as reported
here. All these new $3d$, $4d$, and $5d$-metal substitutions
deserve both an experimental as well as a sophisticated
theoretical treatment, including the investigation of the
influence of disorder and charge doping.


\begin{thebibliography}{38}
\bibitem[{FeC()}]{FeConote}
\bibinfo{note}{{For SrFeCoAs$_2$ the structural data of SrFe$_2$As$_2$ were
  used, since Fe and Co are very similar in size.}}
\end{thebibliography}

\end{document}